\documentclass[sn-basic]{sn-jnl}

\usepackage{graphicx}%
\usepackage{multirow}%
\usepackage{amsmath,amssymb,amsfonts}%
\usepackage{amsthm}%
\usepackage{mathrsfs}%
\usepackage[title]{appendix}%
\usepackage{xcolor}%
\usepackage{textcomp}%
\usepackage{manyfoot}%
\usepackage{booktabs}%
\usepackage{algorithm}%
\usepackage{algorithmicx}%
\usepackage{algpseudocode}%
\usepackage{listings}%
\usepackage{subcaption} 
\usepackage{bm}         

\theoremstyle{thmstyleone}%
\newtheorem{theorem}{Theorem}
\newtheorem{proposition}[theorem]{Proposition}%

\theoremstyle{thmstyletwo}%

\theoremstyle{thmstylethree}%

\begin{document}
\title[A likelihood ratio test for circular multimodality]{A likelihood ratio test for circular multimodality}

\author{Diego Bol\'on\textsuperscript{1}}\email{diego.bolon.rodriguez@ulb.be}

\author{Rosa M. Crujeiras\textsuperscript{1}}\email{rosa.crujeiras@usc.es}

\author{Alberto Rodr\'iguez-Casal\textsuperscript{1}}\email{alberto.rodriguez.casal@usc.es}

\affil{\textsuperscript{1}Galician Center for Mathematical Research and Technology, CITMAga, Universidade de Santiago de Compostela, \ Santiago de Compostela, 15782, Spain}

\abstract{The modes of a statistical population are high frequency points around which most of the probability mass is accumulated. For the particular case of circular densities, we address the problem of testing if, given an observed sample of a random angle, the underlying circular distribution model is multimodal. Our work is motivated by the analysis of migration patterns of birds and the methodological proposal follows a novel approach based on likelihood ratio ideas, combined with critical bandwidths. Theoretical results support the behaviour of the test, whereas simulation examples show its finite sample performance.
}


\maketitle

{\small \textbf{Acknowledgements:} This work has been supported by the Spanish Ministerio de Ciencia, Innovación y Universidades (fellowship FPU20/03960), and by project PID2020-116587GB-I00 supported by Agencia Estatal de Investigación (AEI) from Ministerio de Ciencia, Innovación y Universidades. The Supercomputing Centre of Galicia (CESGA) is acknowledged for providing the computational resources that allowed to run the simulations in Section~\ref{sec:simstudy}. We would also like to thank Jose Ameijeras-Alonso for providing the code of the implementation of the multimodality test by \cite{AmeijeirasAlonso2019}.}


\section{Introduction}\label{sec:intro}

\begin{figure}[!b]
\centering
\includegraphics[width = 0.45\linewidth]{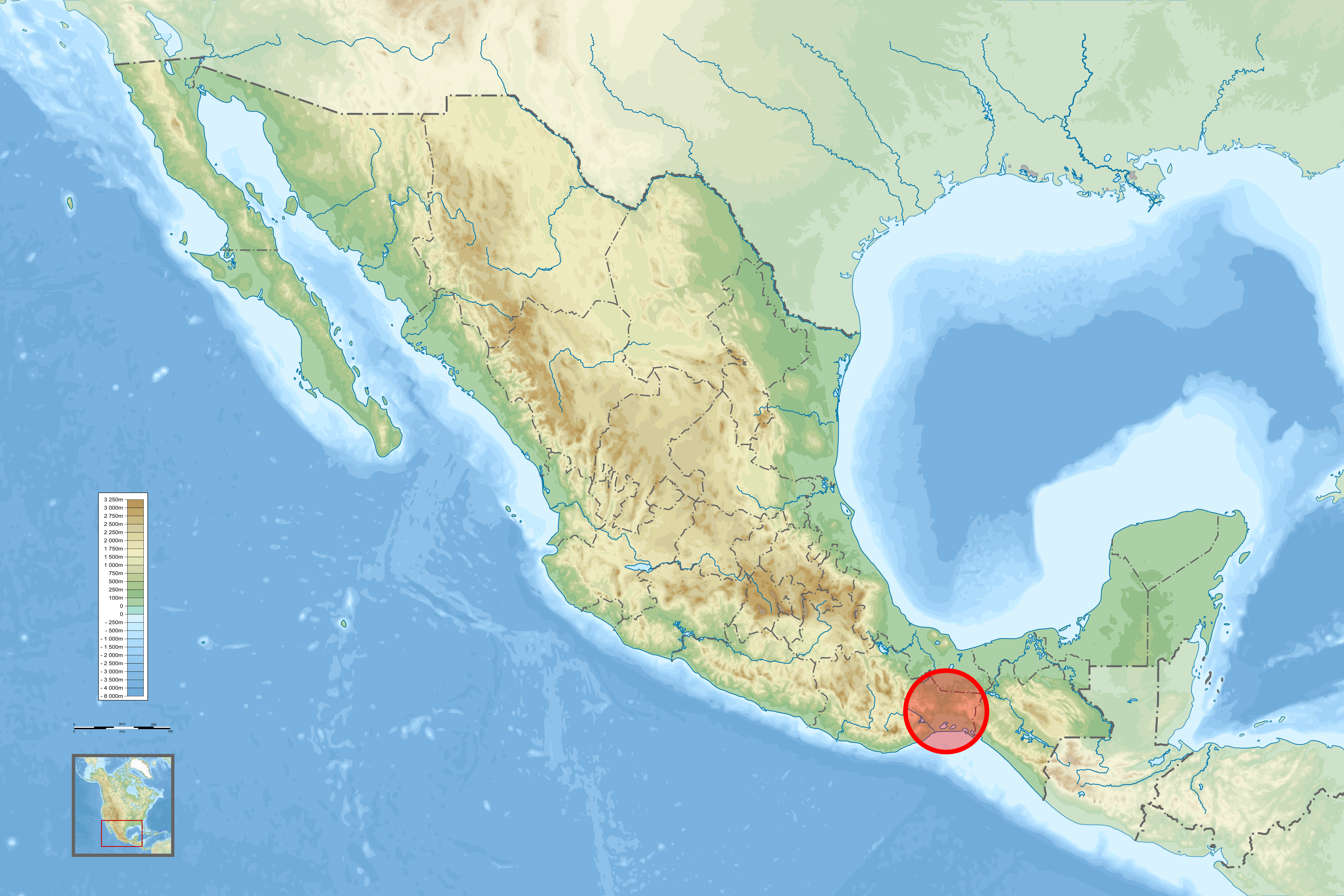}
\hspace{0.05\linewidth}
\includegraphics[width = 0.45\linewidth]{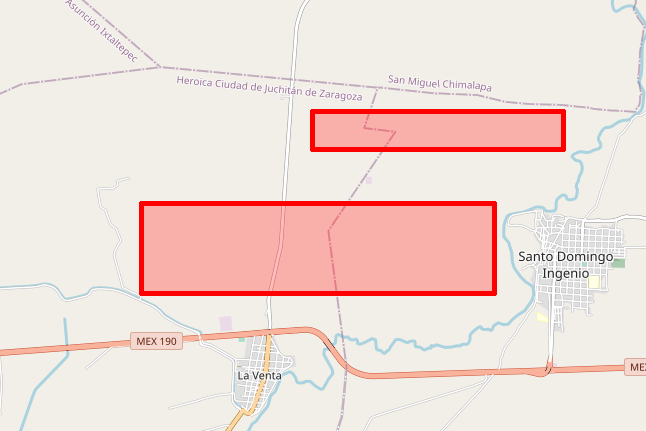}
\vspace{0.1in}
\caption{Left panel: map of Mexico \citep{Mexico} with the location of the observation area marked by a red circle. Right panel: location of the new wind farms erected in the observation area (map retrieved from \citealp{OpenStreetMap}).}
\label{fig:obsarea}
\end{figure}

A mode of a circular random variable is a point where its density function reach a local maximum. The existence of multiple modes in a distribution is intimately related with the presence of different clusters or groups within the population, being modes high frequency points around which most of the probability is accumulated. To motivate the need of assessing multimodality in circular data, that is, to identify whether a (cicular) distribution presents more than a single mode, we introduce the real data example by \cite{CabreraCruz2016}. These authors were interested in determining if the presence of new wind farms had caused a change on the flight routes of migrating raptors. It is a known fact that wind turbines have an impact on bird mortality (see, for instance, \citealp{Maurer2020}). But, do migrating birds actively change their migration routes in order to avoid wind farms? To answer this question, \cite{CabreraCruz2016} recorded flight trajectories in six consecutive autumn migration seasons in the Isthmus of Tehuantepec, Mexico. During that time, two new wind farms were built in the observation area. Figure~\ref{fig:obsarea} shows the location of the observation area within Mexico (letf panel), and the location of the new wind farms in the observation area (right panel). These wind farms lie in the middle of the natural route of many migrating birds that fly from North to South America in order to spend the boreal winter in warmer latitudes. The authors aimed to check if the new constructions had altered the raptors' migration routes and, for that purpose, they compared the directions of the observed trajectories between two different periods: 2009-2011, where there was only one wind farm in the observation zone (pre-construction period), and 2012-2014 where the two new wind farms were erected (post-construction period).

\begin{figure}[t]
\centering
\includegraphics[scale=0.75]{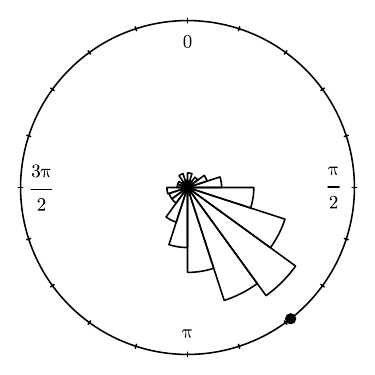}
\includegraphics[scale=0.75]{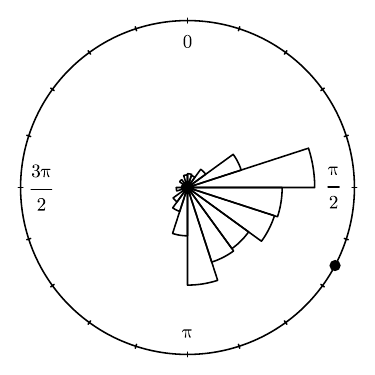}
\caption{Rose diagrams of trajectories directions registered by \cite{CabreraCruz2016}. Left: pre-construction period ($n=1183$). Right: post-construction period ($n=1988$). Angle $0$ corresponds with north, $\pi/2$ with east, $\pi$ with south and $3\pi/2$ with west direction. The mean direction of the sample is represented with a bullet in the circumferences. Circular mean directions were approximately $2.475$ rads. for the pre-contruction period and $2.057$ rads. for the post-contruction period.}
\label{fig:farmsrosediag}
\end{figure}

Figure \ref{fig:farmsrosediag} shows the rose diagrams of observed trajectories during the pre-construction (left panel) and post-construction (right panel) periods, along with their mean directions (marked with bullets in the circumferences). From these simple plots, it is quite clear that there are substantial differences between both distributions, although it may be argued that the output of a rose diagram can be influenced by the bin arc-width considered and visual appreciations are always subjective. \cite{CabreraCruz2016} applied the Watson and Williams test (see \citealp{Watson1956} for details on the procedure) and concluded that there were significant differences between the two samples. It should be noted that the Watson and Williams test is devised to assess differences in mean directions, so any other possible discrepancy is ignored and the conclusion is limited to a change in the mean, but with no further interpretation on other characteristics. For the pre-construction observation, which seem to follow a symmetric and highly concentrated distribution (Figure~\ref{fig:farmsrosediag}, left), the circular mean is a suitable summary of the data distribution. However, this is not the case for the post-construction sample, where observations do not appear to show a unimodal distribution. Hence, the initial analysis fails to provide a complete explanation of how the directions of the migration trajectories changed between the two periods. 

\begin{figure}[t]
\centering
\includegraphics[scale=0.5]{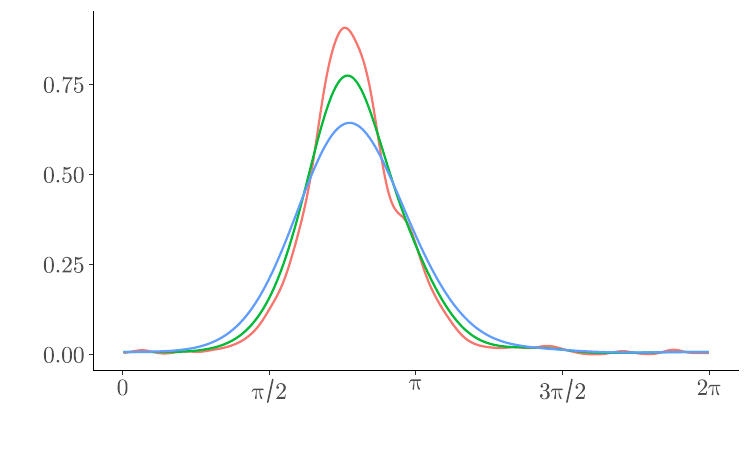}
\includegraphics[scale=0.5]{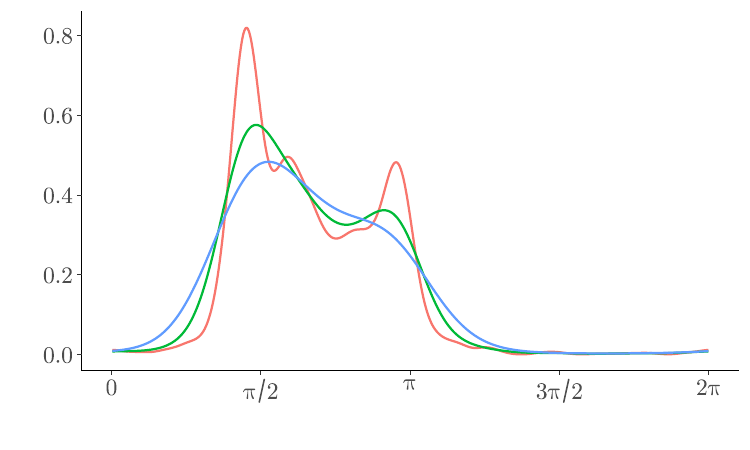}
\caption{Linear representations of circular kernel density estimators of trajectories registered by \cite{CabreraCruz2016}. Left: pre-construction period ($n=1183$). Right: post-construction period ($n=1988$). Angle $0$ corresponds with north, $\pi/2$ with east, $\pi$ with south and $3\pi/2$ with west direction. Each graph shows the kernel density estimator for three different values of the bandwidth: $h = 0.4$ (blue), $h = 0.25$ (green) and $h = 0.1$ (red).}
\label{fig:farmskde}
\end{figure}

Circular kernel density estimators (see equation~\eqref{eq:kdecirc} below) are also a useful tool for graphical representation of circular data. In Figure~\ref{fig:farmskde}, three different kernel density estimators are plotted for the directions of the pre-construction (left panel) and post-construction (right panel) periods. Figure~\ref{fig:farmskde} perfectly illustrates the main flaw of these kinds of estimators: their highly dependence on the value of the bandwidth $h$. This is particularly evident for the post-construction data (right panel), where the kernel density estimators have a very different shape for the three values of $h$ considered. For $h = 0.4$, the estimated density function is unimodal and its mode is located at $\pi/2$. However, for $h = 0.25$, the corresponding estimator has two modes, one at $\pi/2$ and a secondary one at $\pi$. For the smallest bandwidth considered ($h = 0.1$), a new mode arises between $\pi/2$ and $\pi$.

\begin{figure}[t]
\centering
\includegraphics[scale=0.5]{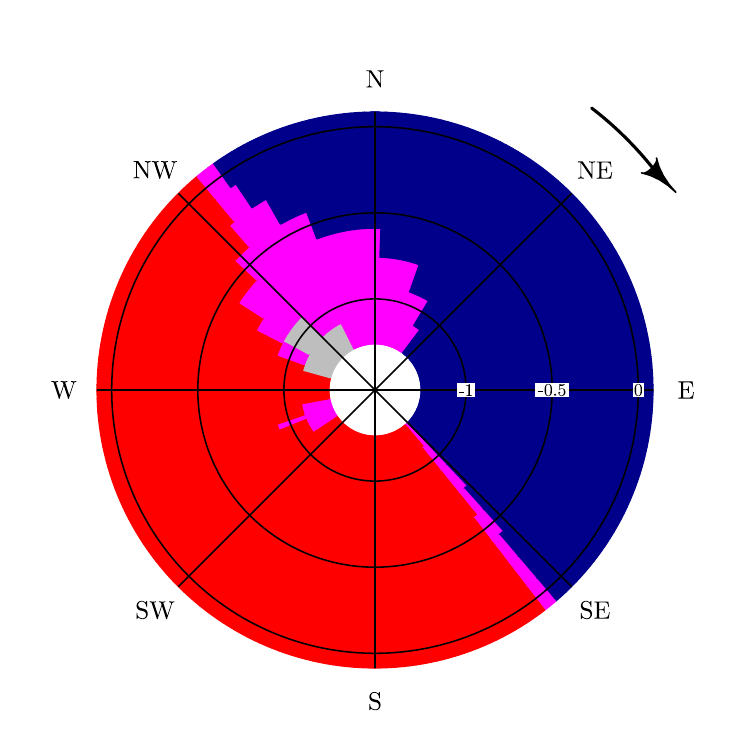}
\includegraphics[scale=0.5]{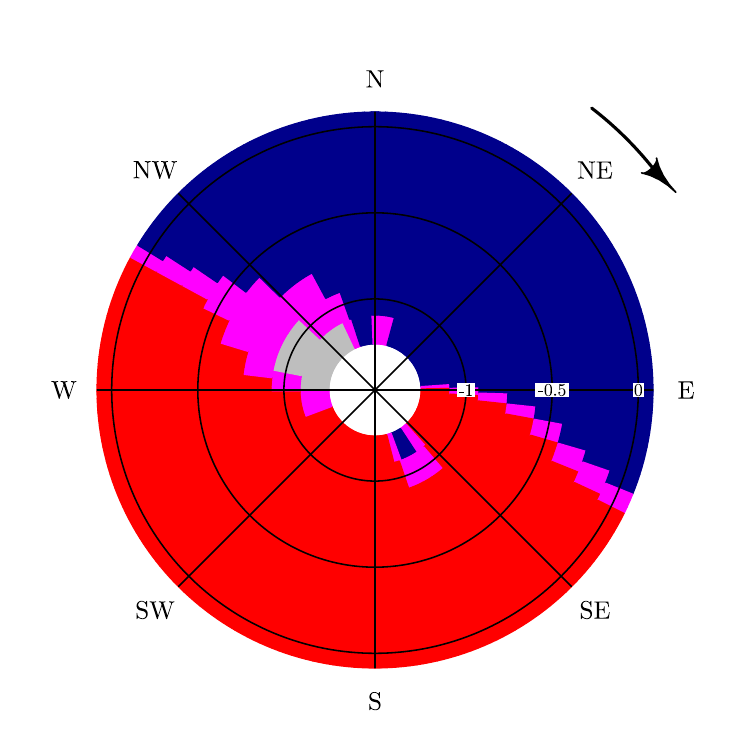}
\caption{CircSiZers for the trajectories directions. Left: pre-construction period. \mbox{($n=1183$)} Right: post-construction period ($n=1988$). Angle $0$ corresponds with north, $\pi/2$ with east, $\pi$ with south and $3\pi/2$ with west direction.}
\label{fig:CircSiZer}
\end{figure}

\cite{Oliveira2014} use circular kernel density estimates as the basis for a more sophisticated exploratory tool: the CircSiZer, which allows to determine the existence of significant modes in a data sample. In Figure \ref{fig:CircSiZer}, the CircSiZers for each period are shown. In a CircSiZer, as explained in \cite{Oliveira2014}, a significant increasing (decreasing) pattern in the estimated density is shown in blue (red), so a blue-red pattern indicates the presence of a high frequency point, that is, a circular mode. Note that, for the pre-construction observations just the single mode observed in the rose diagram is significant. However, for the post-construction samples, there is a principal mode (with some displacement with respect to the mode identified for the pre-construction sample) but a secondary mode is also noticed, although for a limited range of smoothing parameters. This will show that most of the raptors fly eastwards, keeping the wind farms to the south, whereas a small subgroup heads south, avoiding the wind farms by flying west of them (see Figure~\ref{fig:obsarea} for a reference on where the wind farms are located). In view of the CircSiZer results, we may wonder if both migration patterns are unimodal and there is just a displacement trying to avoid the wind farms or if the second group shows a secondary significant mode, showing the existence of a different migration strategy.

A formal approach to assess the changes, ensuring the existence of a single mode in the pre-construction sample and two-modes in the post-construction observations, is possible through the use of multimodality tests. Specifically, a multimodality test is a statistical procedure that checks whether the unknown number of modes is lower or greater than some natural number. More precisely, if $X$ is a random angle with $j$ modes, a multimodality test assesses the hypotheses:
\begin{equation}
\mathcal{H}_0: j \leq k \text{ vs } \mathcal{H}_1: j > k;
\label{eq:testmultimod}
\end{equation}
where $k$ is a given natural number. In particular, the simplest (yet useful) case is the unimodality test:
\begin{equation*}
\mathcal{H}_0: j = 1 \text{ vs } \mathcal{H}_1: j > 1.
\end{equation*}

The problem of testing for multimodality for real-valued random variables has been addressed by different authors. Based on a kernel density estimator, \cite{Silverman1981} proposed the use of the critical bandwidth as a test statistic (whose calibration was corrected by \citealp{Hall2001}), whereas \cite{Mueller1991} and \cite{Cheng1998} considered excess-mass ideas to introduce a testing procedure. \cite{AmeijeirasAlonso2018} combined both views in a procedure that outperforms the existing proposals. For circular random variables, up to the authors' knowledge, the only available tests for multimodality are the ones proposed by \cite{Fisher2001}, which has been shown to present an unsatisfactory behaviour by \cite{AmeijeirasAlonso2019}, extending these authors the ideas of excess-mass combined with critical bandwidth to assess multimodality in the circular setting.

In this work, we aim to introduce a multimodality test for circular data that first, is at least competitive with the existing proposal and outperforms it when this one fails; and second, the methodology is general enough to be directly adapted to other settings (e.g. directional, toroidal or cylindrical data). For that purpose, we will consider the likelihood ratio ideas, with a nonparametric view, using kernel circular density estimators. All these tools will be introduced in subsequent sections.

The manuscript is organized as follows. In Section \ref{sec:background}, some basic ideas for circular data modelling are presented, jointly with the existing proposals for multimodality testing. Section \ref{sec:newproposal} introduces the new multimodality test for circular data, detailing the calculation of the test statistic and the calibration method. In Section \ref{sec:simstudy} the behavior of the new test in practice is analyzed with a simulation study, and its performance is compared with its competitor. The new proposal is applied to the real data example in Section~\ref{sec:data}. Section \ref{sec:conclusion} contains the main conclusions of this paper and some discussion on possible extensions. The paper is acompained with two appendices with supplementary material. Appendix~\ref{sec:A1} contains the proof of the main result of this work. For reproducibility purposes, data and code availability are addressed in Appendix~\ref{sec:A2}. Appendix~\ref{sec:A3} lists all the models considered in the simulation study.
\section{Some background on testing multimodality for circular data}\label{sec:background}
Circular observations are points in the circumference $\mathbb{S}^1 = \{x \in \mathbb R^2 : \|x\| = 1 \}$. The statistical analysis of these types of data has received substantial attention in the literature; see, for instance, \cite{Fisher1993}, \cite{Mardia2000}, \cite{Pewsey2013}, \cite{Ley2017} and \cite{SenGupta2022}. A common strategy is to identify each point of $\mathbb{S}^1$ with an angle between $0$ and $2\pi$. This allows to work with circular data as random angles. A random angle $X$ is a Borel-measurable mapping from a probability space $(\Omega, \mathcal{A}, \mathbb P)$ to the interval~$(0,2 \pi]$:
\begin{equation*}
	X: \Omega \rightarrow (0, 2 \pi].
\end{equation*}

The distribution of an absolutely continuous random angle $X$ is fully determined by its density function $f$ \citep[Sec.~3.2]{Mardia2000}. This function $f$ verifies:
\begin{itemize}
    \item $f(x) \geq 0$ almost everywhere.
    \item $f( x ) = f( x + 2\pi)$ almost everywhere.
    \item
	$\int_{\psi}^{\phi} f( x ) d x = \mathbb{P} (\psi < X \leq \phi)$, for all $\phi, \psi \in (0, 2 \pi]$ such that $\psi \leq \phi$.
\end{itemize}

Once the circular density function is determined, the concept of mode naturally arises. A mode of a random angle $X$ is a point in the interval $(0, 2 \pi]$ where its density function reaches a local maximum. Analogously, an antimode of a random angle is a point in the interval $(0, 2 \pi]$ where its density function reaches a local minimum.

Some of the most important families of circular models include the von Mises, cardioid, wrapped Cauchy and wrappped normal distributions. For a complete review of distribution circular models, see for instance \cite[Sec.~3.5]{Mardia2000} or \cite[Ch.~3]{Fisher1993}. For technical reasons, the parametrization for the wrapped normal distribution used in this paper is slightly different from both references above. Hereinafter, a wrapped normal distribution $\mathrm{WN}(\mu, \sigma^2)$, where $\mu \in (0, 2\pi]$ and $\sigma^2 > 0$, is the circular distribution with density function
\begin{equation*}
	f(x) = \frac{1}{\sqrt{2 \pi \sigma^2}} \sum_{m = -\infty}^{+ \infty} \exp \left( \frac{- (x - \mu + 2 \pi m)^2}{2 \sigma^2} \right), \quad x \in (0, 2\pi].
\end{equation*}

\subsection{Multimodality tests for circular data}

Since the modes are the local maxima of the density function, the multimodality testing problem introduced in \eqref{eq:testmultimod} is closely related with density function estimation. Most multimodality tests ultimately rely on an estimator of the density function, like the kernel density estimator (see, for example, \citealp{Oliveira2012}). Let $X_1, \ldots, X_n$ be an i.i.d.~sample of a random angle $X$. The circular kernel density estimator of $X$ is given by
\begin{equation}
	\label{eq:kdecirc}
	\hat{f}_h (x) = \frac{1}{n} \sum_{i = 1}^n \mathcal{K}_h \left( x - X_i \right),
\end{equation}
where $h \in (0, + \infty)$ is a positive real number, and $\mathcal{K}_h$ is a family of symmetric, centered at zero, circular densities indexed by the parameter $h$. The $\mathcal{K}_h$ are called kernel functions, and $h$ is named bandwidth. A common choice of the kernel function is the wrapped normal density $\mathrm{WN}(0, h^2)$:
\begin{equation}
	\label{eq:wnorm}
	\mathcal{K}_h (x) = \frac{1}{\sqrt{2 \pi h^2}} \sum_{m = -\infty}^{+ \infty} \exp \left( \frac{- (x + 2 \pi m)^2}{2 h^2} \right).
\end{equation}
The shape of the kernel estimator $\hat{f}_h$ introduced in~\eqref{eq:kdecirc} depends heavily on the value of the bandwidth $h$. Focusing specifically on the multimodality problem, if the wrapped normal kernel is chosen, the number of modes of $\hat{f}_h$ is a non decreasing function of $h$, as Figure~\ref{fig:farmskde} illustrates (see \citealp{Huckemann2016} for a proof). This monotonicity property allows to define the critical bandwidth for circular data \citep{AmeijeirasAlonso2018}.

\subsubsection{Critical bandwidth test }
\label{sec:Fisher}
The critical bandwidth for $k$ modes, namely $h_k$, is the smallest value of $h$ such that the number of modes of $\hat{f}_h$ is less or equal to $k$. That is 
\begin{equation}
	\label{eq:critbandcric}
	h_{k} = \inf \{h > 0 \colon \hat{f}_h \text{ has } k \text{ or less modes} \}.
\end{equation}

Although the critical bandwidth has been used as a test statistic for multimodality for real-valued random variables (see \citealp{Silverman1981} and \citealp{Hall2001}), multimodality tests for circular data do not employ the critical bandwidth directly as a test statistic. \cite{AmeijeirasAlonso2017} justifies this difference since the properties of the critical bandwidth are not analogue in both contexts. In the real-valued setting, \cite{Hall2001} show that the null distribution of $h_k$ depends on the curvature of the true density $f$ on its critical points. This makes it difficult to approximate the null distribution unless one assumes that $f$ only has one critical point. Since circular density always have more than one critical point (even unimodal densities), the null distribution of $h_k$ can not be directly approximated for circular data.

Trying to overcome this issue, \cite{Fisher2001} propose a multimodality test for circular data based on the critical bandwidth, but without using it directly as a test statistic. Instead, they approach the problem as a goodness-of-fit test, checking how far is the empirical distribution of the sample from the distribution given by $\hat{f}_{h_k}$. The Watson $U^2$ statistic \citep{Watson1961}, a circular version of the Cramér-von Mises statistic, is chosen for this purpose.

Specifically, the proposal by \cite{Fisher2001} employs the following statistic:
\begin{equation}
	\label{eq:uwatson}
	U^2_n = n^{-1} \int_{-\infty}^{+\infty} \left( F_n(x) - \hat{F}_{h_k} (x) - \int_{-\infty}^{+\infty} \left[ F_n(y) - \hat{F}_{h_k} (y) \right] dF_n(y) \right)^2 dF_n(x);
\end{equation}
where $F_n$ is the empirical distribution function of the sample, $h_k$ is the critical bandwidth for $k$ modes, and $\hat{F}_{h_k} (x) = \int_{0}^{x} \hat{f}_{h_k} (y) d y$. The test rejects the null hypothesis $\mathcal{H}_0: j \leq k$ in \eqref{eq:testmultimod} for large values of $U^2_n$. The null distribution of $U^2_n$ is approximated via smoothed bootstrap, resampling from $\hat{f}_{h_k}$.

The Fisher and Marron's test has shown a poor calibration in practice (see \citealp{AmeijeirasAlonso2017}). A possible explanation of this phenomenon can be found in the different converge rates of $h_k$ and the optimal bandwidth for distribution estimation. For real-valued data, \cite{Mammen1992} showed that, if the underling density of the sample has $k$ modes, then the critical bandwidth $h_k$ is of order $n^{-1/5}$, whereas the optimal bandwidth for distribution estimation is of order $n^{-1/3}$ (see \citealp{Altman1995}). Then the $U^2_n$ statistic from~\eqref{eq:uwatson} uses a kernel estimation of the distribution function with a too large bandwidth that may not correctly fit the data. Consequently, $U^2_n$ may take large values even under the null hypothesis.

\subsubsection{Excess mass test}
\label{sec:Ameijeiras}
The other main tool for testing multimodality is excess mass. This concept, introduced by \cite{Mueller1991}, relies on the intuition that around each mode the density function $f$ has bumps, that is, regions where $f$ takes values over some threshold. So, the excess mass tries to measure these bumps by cutting them from a fixed height.

\begin{figure}[!ht]
	\centering
    \includegraphics{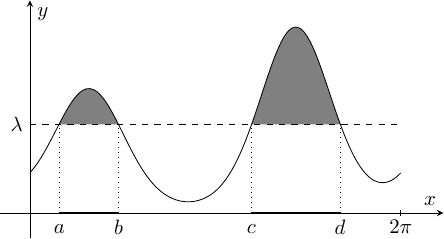} 
	\caption{\label{fig:excesomasa} Linear representation of the excess mass of a circular density function $f$ at level $\lambda$. $E (\lambda)$ is the area of the gray region. In this particular case, the set $L(\lambda)$ consists of two intervals, $[a,b]$ and $[c, d]$, each of them linked to a mode of $f$.} 
\end{figure}

Given $f$ be a circular density function and $\lambda > 0$ a positive real number, the excess mass of $f$ at level $\lambda$ is
\begin{equation*}
	E (\lambda) = \int_{L (\lambda)} f(x) dx - \lambda \mathrm{Vol} \big( L(\lambda) \big) = \mathbb P \big( L(\lambda) \big) - \lambda \mathrm{Vol} \big( L(\lambda) \big),
\end{equation*}
where $L(\lambda) = \{ x \in (0, 2 \pi] : f(x) \geq \lambda\}$ and $\mathrm{Vol} \big( L(\lambda) \big)$ is the Lebesgue measure of the set~$L(\lambda)$. $E (\lambda)$ is measuring the area of the region $\{ (x, y) \in (0, 2 \pi] \times \mathbb R \colon \lambda < y < f(x) \}$, as Figure~\ref{fig:excesomasa} illustrates.

If $f$ is $k$-modal, then $f$ has $k$ bumps, each of them linked to a mode. This means that the set $L(\lambda)$ is a disjoint union of $k+1$ intervals at most: one for each mode, and one more if one of the modes is located near $2\pi$ or $0$ (since, in this case, one of the bumps may extend beyond the interval $(0, 2\pi]$). If this last scenario occurs, two of the $k+1$ disjoint intervals composing $L(\lambda)$ are of the form $(0, a]$ and $[b, 2 \pi]$. The \hbox{$2\pi$-periodicity} of $f$ allows to express the union of these two intervals as a unique interval of the form $[b, a + 2\pi]$, and its probability is
$$
\mathbb P \big( [b, a + 2\pi] \big) = \int_{b}^{a + 2 \pi} f(x) d x.
$$
Then, if $f$ at least $k$ modes, one can express excess mass as
\begin{equation}
	E (\lambda) = \sup_{C_1, \ldots, C_k} \left\lbrace \sum_{c = 1}^{k} \mathbb P (C_c) - \lambda \mathrm{Vol} ( C_{c} ) \right\rbrace;
	\label{eq:exckmodascirc}
\end{equation}
where the supremum is taken over all families of $k$ pairwise disjoint intervals of~$(0, 4\pi]$, $\{C_1, \ldots, C_k \}$. The left side of \eqref{eq:exckmodascirc} can be estimated empirically, replacing the probability of $f$ by the probability of the empirical distribution. Then,
\begin{equation*}
	E_{n, k} (\lambda) = \sup_{C_1, \ldots, C_k} \left\lbrace \sum_{c = 1}^{k} \mathbb P_n (C_c) - \lambda \mathrm{Vol} ( C_{c} ) \right\rbrace
\end{equation*}
is a natural estimator of the excess mass, where $\mathbb P_n$ is the probability of the empirical distribution and, as before, the supremum is taken over all families of $k$ pairwise disjoint intervals of~$(0, 4\pi]$, $\{C_1, \ldots, C_k \}$.

However, the quantity $E_{n, k} (\lambda)$ will be a consistent estimator of $E (\lambda)$ if and only if $f$ has at most $k$ modes. Therefore, the difference $\big[ E_{n, k+1} (\lambda) - E_{n, k} (\lambda) \big]$ should somehow measure how far the null hypothesis $\mathcal{H}_0: j \leq k$ is from being true. So the quantity
\begin{equation*}
	\Delta_{n, k+1} = \max_{\lambda > 0} \big[ E_{n, k+1} (\lambda) - E_{n, k} (\lambda) \big]
\end{equation*}
seems to be a good proposal for testing multimodality of circular data. The null hypothesis will be rejected for large values of $\Delta_{n, k+1}$.

\cite{AmeijeirasAlonso2019} propose a multimodality test for circular data using this test statistic. Their idea is to approximate the null distribution of $\Delta_{n, k+1}$ resampling from a slightly modified version of $\hat{f}_{h_k}$. This modification is inspired by the asymptotic properties of $\Delta_{n, k+1}$ in the real-valued case.

Assuming that the results by \cite{Cheng1998} also hold for circular data, the null distribution of $\Delta_{n, k+1}$ depends on the curvature of $f$ in its critical points. Specifically, if $f$ has $k$ modes and $x_1, \ldots, x_{2k} \in (0, 2 \pi]$ are the critical points of $f$, the asymptotic distribution of $\Delta_{n, k+1}$ only depends on the values
\begin{equation*}
	d_l = \frac{|f'' (x_l)|}{{f}^3 (x_l)}, \quad l \in \{ 1, \ldots, 2k \}.
\end{equation*} 
\cite{AmeijeirasAlonso2019} suggest to estimate these $d_l$ values by kernel estimation methods. Then, they modify the density function $\hat{f}_{h_k}$ so its $d_l$ values are equal to their estimates, and finally, resampling is performed from this modified version of $\hat{f}_{h_k}$. The proposed estimators for the $d_l$ values are
\begin{equation*}
	\hat{d}_l =  \frac{|\hat{f}_{h'}'' (\hat{x}_l)|}{\hat{f}_{h_j}^3 (\hat{x}_l)}, \quad l = 1, \ldots, 2k
\end{equation*}
where $\hat{f}_{h_k}$ and $\hat{f}''_{h'}$ are the kernel estimators of $f$ and $f''$, $h_k$ is the critical bandwidth for $k$ modes, $h'$ is the bandwidth which minimizes the asymptotic mean integrated squared error of $\hat{f}_{h'}''$ assuming that $f$ is a mixture of von Mises, and $\hat{x}_1, \ldots, \hat{x}_{2k}$ are the critical points of $\hat{f}_{h_k}$.

The density $\hat{f}_{h_k}$ should be slightly modified to obtain a new density, $g_0$, such that its $d_l$ values are equal to their estimators. This new density $g_0$ verifies that $g_0 (x) = \hat{f}_{h_k} (x)$ except on small neighborhoods around the $2k$ critical points. In those parts small changes are made so that the position of the modes and antimodes of $\hat{f}_{h_k}$ and $g_0$ match, and 
\begin{equation*}
	\hat{d_l} = \frac{|g_0'' (\hat{x}_l)|}{g_0^3 (\hat{x}_l)}, \quad \quad \hat{f}_{h_k} ( \hat{x}_l ) = g_0 (\hat{x}_l), 
\end{equation*}
for $l = 1, \ldots, 2k$. Finally, \cite{AmeijeirasAlonso2019} approximate the null distribution of $\Delta_{n, k+1}$ resampling from this new density $g_0$. A detailed explanation about how the density $\hat{f}_{h_k}$ is modified can be found in \cite{AmeijeirasAlonso2019}.

\section{A new circular multimodality test}\label{sec:newproposal}
In this section, a new multimodality test for circular data is introduced. This new proposal is based on the likelihood ratio test ideas translated to a non parametric context. First, a non parametric analogue of the likelihood function is required, which is achieved via the kernel density estimator.

Let $X_1, \ldots, X_n$ be an i.i.d.~sample of a random angle $X$ with density $f$. The \textit{pseudo-likelihood function} of the sample is the real function
\begin{equation}
	\label{eq:pslikehood}
	\mathcal{L} (h) = \prod_{i = 1}^{n} \hat{f}_h ( X_i ),
\end{equation}
where $h > 0$, and $\hat{f}_h$ is the kernel density estimator defined in \eqref{eq:kdecirc}. The function $\mathcal{L}$ can be interpreted as a parametric likelihood function where the assumed parametric density family is indexed by $h > 0$, that is, $\{ \hat{f}_h \colon h > 0 \}$. However, this family is not independent of the data and it is directly specified by the sample.

The new multimodality test follows the structure of a parametric likelihood ratio test. So, the two hypotheses in \eqref{eq:testmultimod} should be translated into a division of the parameter space $\Theta = (0, +\infty)$ in order to define the test statistic. This is performed relying on the kernel density estimator again. Then, the values of $h$ with a kernel density estimator with $k$ or less modes are linked with the null hypothesis, and the remaining values to the alternative hypothesis.

This division of the parameter space can be simplified taking into account the critical bandwidth $h_k$ in \eqref{eq:critbandcric}. By definition, $h_k$ is the limit between kernel density estimators with more and less than $k$ modes. If the bandwidth $h$ is smaller than $h_k$, then $\hat{f}_h$ has more than $k$ modes, and if $h$ is larger than $h_k$, then the number of modes of $\hat{f}_h$ is at most $k$. This division completely agrees with the separation between the null and alternative hypotheses in \eqref{eq:testmultimod}. Then, the parameter space $\Theta$ can be split by the critical bandwidth as follows
\begin{align*}
	\Theta_0 &= \{ h > 0 \colon \hat{f}_h \text{ has } k \text{ or less modes}\} = [h_k, + \infty); \\
	\Theta_1 &= \{ h > 0 \colon \hat{f}_h \text{ has more than } k \text{ modes}\} = (0, h_k).
\end{align*}

Based on these ideas, the test statistic for addressing \eqref{eq:testmultimod} is given by:
\begin{equation}
	\label{eq:fakeDk}
	D_k = 2 \left[ \sup_{h > 0} \ell ( h ) - \sup_{ h \geq h_k} \ell ( h ) \right],
\end{equation}
where $\ell ( h ) = \log \mathcal{L} ( h )$. Similarly to parametric likelihood ratio tests, large values of $D_k$ lead to reject the null hypothesis. 

However, the $D_k$ statistic introduced in \eqref{eq:fakeDk} is not well defined. It can be seen that $\lim_{h \rightarrow 0} \mathcal{L} (h) = +\infty$. Therefore $\mathcal{L}$ is not upper bounded and $\sup_{h > 0} \mathcal{L} ( h )$ is not a real number. One way to solve this issue is to redefine the function $\mathcal{L}$ given by \eqref{eq:pslikehood} using cross-validation. Let $\hat{f}^{-i}_h$ be the kernel density estimator in \eqref{eq:kdecirc} without the $i$-th observation:
\begin{equation*}
	\hat{f}^{-i}_h (x) = \frac{1}{n - 1} \sum_{m = 1, m \neq i}^{n} \mathcal{K}_{h} \left( x - X_m \right).
\end{equation*}
Then, the \textit{cross-validation pseudo-likelihood} of the sample $X_1, \ldots, X_n$ is defined as
\begin{equation}
	\label{eq:CVpslikehood}
	\mathcal{L}_{CV}(h) = \prod_{i = 1}^{n} \hat{f}^{-i}_h ( X_i ), \quad h > 0.
\end{equation}
The new function $\mathcal{L}_{CV}$ is upper bounded, as Proposition~\ref{prop:maxCVcirc} guarantees. The proof of this result can be found in Appendix~\ref{sec:A1}.\\

\begin{proposition}
	\label{prop:maxCVcirc}
	Let $X_1, \ldots, X_n$ be an i.i.d.~sample of an absolutely continuous random angle $X$. Let $\mathcal{L}_{CV}$ be the cross-validation pseudo-likelihood function defined in \eqref{eq:CVpslikehood}, where $\mathcal{K}_h$ is the wrapped normal density $\mathrm{WN}(0, h^2)$ in \eqref{eq:wnorm}. Then, with probability~$1$, the function $\mathcal{L}_{CV}(h)$ is upper bounded in $(0, + \infty)$.
\end{proposition}

\vspace*{0.5cm}
Then, the test statistic 
\begin{equation}
	\label{eq:Dk}
	D_k = 2 \left[ \max_{h > 0} \ell_{CV} ( h ) - \max_{ h \geq h_k} \ell_{CV} ( h ) \right],
\end{equation}
where $\ell_{CV} ( h ) = \log \mathcal{L}_{CV} ( h )$, is well defined by Proposition \ref{prop:maxCVcirc}. As before, large values of $D_k$ reject the null hypothesis. It should be noted that $D_k$ will be null whenever $\max_{h > 0} \ell_{CV} ( h ) = \max_{ h \geq h_k} \ell_{CV} ( h )$. Consequently, $\mathbb{P} (D_k = 0) > 0$ may hold for some distributions. 

\begin{table}[t]
	\caption{\label{tab:nullDk}
		Estimates of $\mathbb{P} (D_1 = 0)$ for Models 1--5.}
	\centering
	\begin{tabular}[t]{cc}
		\toprule
		Distribution & $\mathbb{P}_{\mathcal{M}} (D_1 = 0)$ \\
		\midrule
		Model 1 & 0.380 \\
        Model 2 & 0.244 \\
        Model 3 & 0.331 \\
        Model 4 & 0.291 \\
        Model 5 & 0.037 \\
		\bottomrule
	\end{tabular}
\end{table}

\begin{figure}[t]
\centering
\includegraphics[scale=0.5]{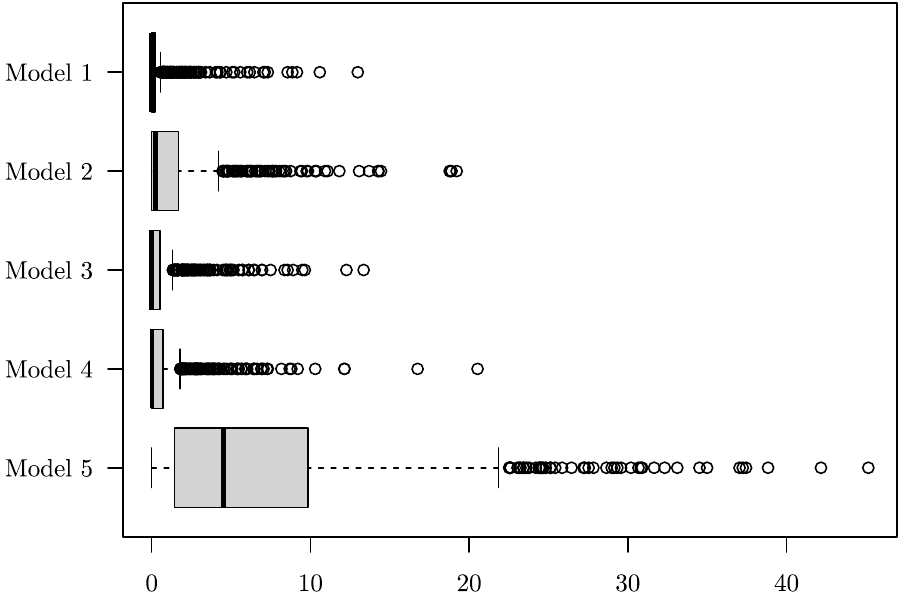}
\vspace{0.1in}
\caption{Boxplots of $1000$ replicates of the test statistic $D_1$ (conditioned on $D_1 > 0$) for five unimodal distributions. Each replicate was computed from a sample of size $n = 1000$. See Appendix~\ref{sec:A3} for a definition of the five models.}
\label{fig:nullDk}
\end{figure}

For test calibration, one might think of deriving the (asymptotic) distribution of $D_k$ under the null hypothesis. However, our simulation results indicate that the null distribution of $D_k$ is not distribution-free, that is, the test statistic $D_k$ has a different distribution for each circular population under $\mathcal{H}_0: j \leq k$. We simulated $\mathcal{M} = 1000$ samples of size $n = 1000$ from five unimodal circular distributions (Models 1 to 5, see Appendix~\ref{sec:A3} for a definition of these distributions), and the test statistic $D_1$ was computed for each sample. Thus, we obtained $\mathcal{M} = 1000$ replicates of $D_1$, $D_1^1, \ldots, D_1^{1000}$, that were used to approximate the null distribution of the test statistic. For example, the probability of $\mathbb{P} (D_1 = 0)$ was estimated by the proportion of replicates equal to zero, i.e.:
$$\mathbb{P}_{\mathcal{M}} (D_1 = 0) = \frac{1}{\mathcal{M}} \sum_{m = 1}^{\mathcal{M}} \mathbb{I} (D_1^{m} = 0).$$
Table~\ref{tab:nullDk} shows the estimates of $\mathbb{P} (D_1 = 0)$ for the five distributions considered. They are significantly different for the five models. Moreover, in Figure~\ref{fig:nullDk}, the boxplots of the non-null replicates of $D_1$ are plotted. There exist remarkable differences between the distribution of $D_1$ (conditioned on $D_1 > 0$) under the five models. These differences are especially noticeable on the right tails, which are extremely important for test calibration since the null hypothesis is rejected for large values of $D_1$.

Therefore, our simulation results raise some skepticism about the practical utility of determining the null distribution of $D_k$. Instead, a fully data-driven calibration method is proposed.

\subsection{Calibration procedure}
Smoothed bootstrap is employed for test calibration. Following the ideas from \cite{AmeijeirasAlonso2019}, resamples are simulated from the circular kernel density estimator with the critical bandwidth, namely $\hat{f}_{h_k}$. Replicates of $D_k$ are calculated from these resamples, and the p-value of the test is approximated by the proportion of replicates larger than the observed value of the statistic.

Finally, as a summary, an outline of the implementation of the new multimodality test with significance level $\alpha \in (0, 1)$ for circular data is provided.
\begin{enumerate}
	\item With the original sample of random angles $X_1, \ldots, X_n$, calculate the critical bandwidth $$h_{k} = \min \{h > 0 \colon \hat{f}_h \text{ has } k \text{ or less modes} \}$$ and the bandwidths which maximize the cross-validation pseudo-likelihood under the null and alternative hypothesis:
	$$\ell_{CV}(h_{\mathrm{max}}) = \max_{h > 0} \{\ell_{CV} (h)\}; \quad \ell_{CV}(h_{\mathcal{H}_0}) = \max_{h \geq h_{k}} \{\ell_{CV} (h)\}.$$
	Compute the test statistic as:
	\begin{equation*}
		D_k = 2 \left[ \ell_{CV}(h_{\mathrm{max}}) - \ell_{CV}(h_{\mathcal{H}_0}) \right].
	\end{equation*}
	\item The resample $X_1^{\ast}, \ldots, X_n^{\ast}$ is simulated from the smoothed density $\hat{f}_{h_{k}}$. One replicate of the test statistic, denoted by $D_k^{\ast}$, is calculated from the resample.
	\item Repeat step 2 $B$ times. Then one will obtain $B$ replicates of the test statistic: $D_k^{\ast, 1}, D_k^{\ast, 2}, \ldots, D_k^{\ast, B}$.
	\item The null hypothesis ($\mathcal{H}_0\colon f$ has $k$ or less modes) is rejected if the approximated p-value $$\frac{1}{B} \sum_{b = 1}^{B} \mathbb{I} (D_k^{\ast, b} > D_k)$$ is smaller than $\alpha$.
\end{enumerate}

\subsection{The $h_{\mathrm{max}}$ bandwidth}\label{subsec:hmax}

The pseudo-likelihood in \eqref{eq:CVpslikehood} function can be also used for bandwidth selection for kernel density estimation. Given the circular sample $X_1, \ldots, X_n$, one could choose as bandwidth the value $h_{\mathrm{max}}$ verifying:
\begin{equation*}
	\mathcal{L}_{CV} (h_{\mathrm{max}}) = \max_{h > 0} \mathcal{L}_{CV} (h);
\end{equation*}
which exists since $\mathcal{L}_{CV}$ is upper bounded.

The main properties of this bandwidth were analyzed by \cite{Hall1987a}, who proved that $h_{\mathrm{max}}$ is an asymptotically global optimal bandwidth. This means that the kernel density estimator $\hat{f}_{h_{\mathrm{max}}}$ is an asymptotically consistent estimator of the underlying density of the population $f$, if $f$ is assumed to be bounded away from zero.
It is expected that this optimality property of the $h_{\mathrm{max}}$ bandwidth translates into a good behavior of the new multimodality test.

Furthermore, the conditions needed for asymptotic convergence of $\hat{f}_{h_{\mathrm{max}}}$ show a possible flaw of the multimodality test based on pseudo--likelihood. The density function $f$ is required to be bounded away from zero for guaranteeing the asymptotic convergence of $\hat{f}_{h_{\mathrm{max}}}$, so the test may have problems dealing with densities vanishing at a point or in an arc of positive length. The simulation study in Section~\ref{sec:simstudy} includes some densities of this type to check whether or not this is a real issue of the test in practice.

\section{Simulation study}\label{sec:simstudy}

A simulation study was performed to check the calibration and power of the new test, and to compare its behavior in practice with the existing multimodality tests for circular data. For this purpose, $\mathcal{M} = 1000$ samples of size $n$ (with $n = 100$, $n = 500$, and $n = 1000$) were simulated from various circular distributions. The new test with significance level $\alpha$ was applied to every sample, and the proportion of rejections was calculated. Fifteen different circular distributions have been considered: five with one mode, five with two modes, and five with three modes. Details of these fifteen distributions can be found in Appendix~\ref{sec:A3}. The three most common significance levels were chosen for the simulation study: $\alpha = 0.01$, $\alpha = 0.05$ and $\alpha = 0.1$. In all cases, $B = 500$ resamples were employed for approximating the p-value.

The simulation study addressed the multimodality test introduced in \eqref{eq:testmultimod} for two values of $k$: $k = 1$ (unimodality vs. multimodality) and $k = 2$ (having at most two modes vs. having more than two modes). We discuss the obtained results separately for these two cases.

\subsection{Testing unimodality} \label{sec:simunimod}

We start regarding the behavior in practice of the new test when testing unimodality vs. multimodality ($k = 1)$. As reported in \cite{AmeijeirasAlonso2017}, the test proposed by \cite{Fisher2001} suffers from poor calibration. Hence, in our simulations, we only include the multimodality test by \cite{AmeijeirasAlonso2019} for comparison. The obtained results can be found in Tables \ref{tab:resultacc} and \ref{tab:resultpow}. The new proposal is referred as likelihood test in both tables.

\begin{table}[tbp]
	\caption{\label{tab:resultacc}
		Proportions of rejections for the test based on pseudo-likelihood \eqref{eq:Dk} and the test by \cite{AmeijeirasAlonso2019} with significance level 1\%, 5\% and 10\% calculated from $\mathcal{M} = 1000$ samples. $B = 500$ resamples were employed for test calibration. Each row represents a different combination of sampling distribution and sample size. All the considered distributions are unimodal. See Appendix~\ref{sec:A3} for a definition of the models.}
	\centering
	\begin{tabular}[t]{llllllll}
		\toprule
		&  & \multicolumn{3}{c}{Likelihood test} & \multicolumn{3}{c}{Ameijeiras et al.~test}\\
		\cmidrule(l{3pt}r{3pt}){3-5} \cmidrule(l{3pt}r{3pt}){6-8}
		\multirow{-2}{*}{Model} & \multirow{-2}{*}{Size} & 1\% & 5\% & 10\% & 1\% & 5\% & 10\%\\
		\midrule
		& $n = 100$ & 0.001 & 0.013 & 0.037 & 0.007 & 0.036 & 0.067\\
		
		& $n = 500$ & 0.005 & 0.032 & 0.060 & 0.006 & 0.029 & 0.073\\
		
		\multirow{-3}{*}{\raggedright M1} & $n = 1000$ & 0.003 & 0.027 & 0.058 & 0.003 & 0.039 & 0.072\\
		\cmidrule{1-8}
		& $n = 100$ & 0.005 & 0.036 & 0.080 & 0.012 & 0.049 & 0.113\\
		
		& $n = 500$ & 0.011 & 0.054 & 0.096 & 0.013 & 0.049 & 0.105\\
		
		\multirow{-3}{*}{\raggedright M2} & $n = 1000$ & 0.011 & 0.056 & 0.110 & 0.009 & 0.052 & 0.109\\
		\cmidrule{1-8}
		& $n = 100$ & 0.003 & 0.029 & 0.059 & 0.008 & 0.032 & 0.077\\
		
		& $n = 500$ & 0.013 & 0.046 & 0.087 & 0.004 & 0.036 & 0.079\\
		
		\multirow{-3}{*}{\raggedright M3} & $n = 1000$ & 0.006 & 0.033 & 0.070 & 0.004 & 0.038 & 0.087\\
		\cmidrule{1-8}
		& $n = 100$ & 0.011 & 0.048 & 0.075 & 0.003 & 0.031 & 0.075\\
		
		& $n = 500$ & 0.009 & 0.039 & 0.081 & 0.005 & 0.023 & 0.069\\
		
		\multirow{-3}{*}{\raggedright M4} & $n = 1000$ & 0.006 & 0.034 & 0.057 & 0.005 & 0.035 & 0.072\\
		\cmidrule{1-8}
		& $n = 100$ & 0.011 & 0.063 & 0.133 & 0.011 & 0.039 & 0.087\\
		
		& $n = 500$ & 0.030 & 0.138 & 0.232 & 0.006 & 0.034 & 0.082\\
		
		\multirow{-3}{*}{\raggedright M5} & $n = 1000$ & 0.052 & 0.155 & 0.263 & 0.007 & 0.038 & 0.085\\
		\bottomrule
	\end{tabular}
\end{table}

Table \ref{tab:resultacc} shows the calibration results. For Models 1 to 4, supported in the whole circumference, the proportions of rejections of the likelihood test are similar to or smaller than the significance level for every sample size. The test is notably conservative in Models 1, 3 and 4, with proportions of rejections much smaller than the significance levels. Moreover, in these three cases increasing the sample size from $n = 500$ to $n = 1000$ results in a reduction of the proportions of rejections. This phenomenon may seem surprising at first, but it turns out to be a direct consequence of the nature of the test.

The type I error of likelihood ratio tests with composite null hypothesis is not equal for every distribution under the null hypothesis. These types of tests only achieve a type I error equal to the significance level for scenarios in the boundary between the null and alternative hypothesis. The new multimodality test is based on likelihood ratio tests, so one expects to obtain proportions of rejections much smaller than the significance levels for distributions that are clearly unimodal, such as Models 1 and 4. On the contrary, the test should reach proportions of rejections similar to the significance levels for distributions that are close to be multimodal. This seems to be the case for Model 2, where the test achieves proportions of rejections similar to the significance levels for large sample sizes ($n = 500$ and $n = 1000$).

Finally, the new test shows proportions of rejections larger than the significance levels for Model 5. This confirms that the test does not perform as well as it should when the density function vanishes in a circular sector with positive measure (which is the case for this model), as it was suggested in Section \ref{subsec:hmax}.

The test by \cite{AmeijeirasAlonso2019} obtains proportions of rejections similar to or smaller than the significance levels in every considered scenario. The test exhibits a conservative behavior in Models 1 and 4, but obtains larger proportions of rejections than the new proposal in both cases. Finally, for Model 5 the proportions of rejections are smaller than the significance levels for all sample sizes. This seems to point out that this test has no problem with densities vanishing in an arc of positive length, unlike the likelihood ratio test.

\begin{table}[tbp]
	\caption{\label{tab:resultpow}
		Proportions of rejections for the test based on pseudo-likelihood \eqref{eq:Dk} and the test by \cite{AmeijeirasAlonso2019} with significance level 1\%, 5\% and 10\% calculated from $\mathcal{M} = 1000$ samples. $B = 500$ resamples were employed for test calibration. Each row represents a different combination of sampling distribution and sample size. All the considered distributions are bimodal. See Appendix~\ref{sec:A3} for a definition of the models.}
	\centering
	\begin{tabular}[t]{llllllll}
		\toprule
		&  & \multicolumn{3}{c}{Likelihood test} & \multicolumn{3}{c}{Ameijeiras et al.~test}\\
		\cmidrule(l{3pt}r{3pt}){3-5} \cmidrule(l{3pt}r{3pt}){6-8}
		\multirow{-2}{*}{Model} & \multirow{-2}{*}{Size} & 1\% & 5\% & 10\% & 1\% & 5\% & 10\%\\
		\midrule
		& $n = 100$ & 0.143 & 0.290 & 0.395 & 0.096 & 0.235 & 0.352\\
		
		& $n = 500$ & 0.423 & 0.720 & 0.846 & 0.486 & 0.711 & 0.823\\
		
		\multirow{-3}{*}{\raggedright M6} & $n = 1000$ & 0.963 & 0.993 & 0.999 & 0.819 & 0.928 & 0.960\\
		\cmidrule{1-8}
		& $n = 100$ & 0.034 & 0.083 & 0.129 & 0.020 & 0.091 & 0.160\\
		
		& $n = 500$ & 0.118 & 0.290 & 0.384 & 0.070 & 0.217 & 0.314\\
		
		\multirow{-3}{*}{\raggedright M7} & $n = 1000$ & 0.231 & 0.422 & 0.542 & 0.118 & 0.292 & 0.398\\
		\cmidrule{1-8}
		& $n = 100$ & 0.021 & 0.072 & 0.140 & 0.011 & 0.069 & 0.142\\
		
		& $n = 500$ & 0.034 & 0.135 & 0.217 & 0.021 & 0.084 & 0.142\\
		
		\multirow{-3}{*}{\raggedright M8} & $n = 1000$ & 0.052 & 0.184 & 0.278 & 0.020 & 0.076 & 0.158\\
		\cmidrule{1-8}
		& $n = 100$ & 0.081 & 0.208 & 0.312 & 0.003 & 0.037 & 0.079\\
		
		& $n = 500$ & 0.317 & 0.513 & 0.616 & 0.006 & 0.041 & 0.079\\
		
		\multirow{-3}{*}{\raggedright M9} & $n = 1000$ & 0.556 & 0.683 & 0.749 & 0.006 & 0.042 & 0.079\\
		\cmidrule{1-8}
		& $n = 100$ & 0.154 & 0.379 & 0.530 & 0.006 & 0.032 & 0.080\\
		
		& $n = 500$ & 0.820 & 0.903 & 0.936 & 0.015 & 0.097 & 0.168\\
		
		\multirow{-3}{*}{\raggedright M10} & $n = 1000$ & 0.957 & 0.980 & 0.989 & 0.071 & 0.260 & 0.391\\
		\bottomrule
	\end{tabular}
\end{table}

The power behavior results are shown in Table \ref{tab:resultpow}. The new test detects the alternative hypothesis for all scenarios. Furthermore, the results are coherent with the shape of the sampling distributions. For example, the power of the test increases as the distance between the two modes increases (Model 6 vs.~Model 7). The power also rises with the excess mass of the secondary mode (Model 9 vs.~Model 10). For Model~8 the proportions of rejections are much smaller, but always above the significance level. Additionally, the proportions of rejections grows as the sample size gets larger in all the scenarios.

Comparing the results of both tests, the new test is clearly more powerful than the proposal by \cite{AmeijeirasAlonso2019}, that only achieves proportions of rejections similar to the ones obtained by the new test under certain scenarios with small sample size (Models 7 and 8 with $n = 100$). For most cases, the likelihood-ratio test obtains much larger proportions of rejections than its competitor. This is specially remarkable for distributions with a secondary mode with small excess mass, such as Models 9 and 10, since the proposal by \cite{AmeijeirasAlonso2019} relies on excess mass for testing unimodality. Moreover, this test fails to detect multimodality in Model 9 due to the extremely small excess mass of the secondary mode. For this distribution, all the proportions of rejections obtained by Ameijeiras' test are smaller than the significance levels.

\subsection{Testing for at most two modes} \label{sec:simbimod}

The results obtained when testing for at most two modes vs. more than two modes \mbox{($k = 2$)} are collected in Tables~\ref{tab:resultbimnull} and~\ref{tab:resultbimpow}. In Section~\ref{sec:simunimod}, it has been shown that the new test outperforms the proposal by \cite{AmeijeirasAlonso2018} in terms of power. Therefore, we only include the results of the new test from now on.

Table~\ref{tab:resultbimnull} contains the calibration results. For the unimodal distributions (Models 1 to 4), the test obtains proportions of rejections that are notably smaller than the significance levels for all sample sizes. The behavior of the new test is very different for Model 5. In this case, the test suffers from the same problem of calibration reported in Section~\ref{sec:simunimod}, and the proportion of rejections are larger than the signification levels even for the largest sample size. The new procedure still has problems with densities that are not bounded away from zero when testing bimodality.

\begin{table}[!t]
\centering
\caption{\label{tab:resultbimnull}
		Proportions of rejections for the test based on pseudo-likelihood \eqref{eq:Dk} with significance level 1\%, 5\% and 10\% when testing bimodality. All proportions of rejections were calculated from $\mathcal{M} = 1000$ samples. $B = 500$ resamples were employed for test calibration. Each row represents a different combination of sampling distribution and sample size. All the distributions considered had 2 or less modes. See Appendix~\ref{sec:A3} for a definition of the models.}
\begin{tabular}{lllll|lllll}
\toprule
Model & Size & 1\% & 5\% & 10\% & Model & Size & 1\% & 5\% & 10\%\\
\midrule
 & $n = 100$ & 0.002 & 0.014 & 0.038 &  & $n = 100$ & 0.001 & 0.051 & 0.138\\
 & $n = 500$ & 0.005 & 0.025 & 0.041 &  & $n = 500$ & 0.004 & 0.013 & 0.052\\
\multirow{-3}{*}{\raggedright M1} & $n = 1000$ & 0.004 & 0.019 & 0.043 & \multirow{-3}{*}{\raggedright M6} & $n = 1000$ & 0.006 & 0.020 & 0.047\\
\cmidrule{1-10}
 & $n = 100$ & 0.004 & 0.025 & 0.056 &  & $n = 100$ & 0.002 & 0.016 & 0.050\\
 & $n = 500$ & 0.006 & 0.028 & 0.059 &  & $n = 500$ & 0.005 & 0.032 & 0.064\\
\multirow{-3}{*}{\raggedright M2} & $n = 1000$ & 0.008 & 0.027 & 0.059 & \multirow{-3}{*}{\raggedright M7} & $n = 1000$ & 0.004 & 0.037 & 0.071\\
\cmidrule{1-10}
 & $n = 100$ & 0.003 & 0.026 & 0.051 &  & $n = 100$ & 0.010 & 0.031 & 0.058\\
 & $n = 500$ & 0.005 & 0.038 & 0.074 &  & $n = 500$ & 0.008 & 0.046 & 0.092\\
\multirow{-3}{*}{\raggedright M3} & $n = 1000$ & 0.005 & 0.023 & 0.052 & \multirow{-3}{*}{\raggedright M8} & $n = 1000$ & 0.021 & 0.055 & 0.107\\
\cmidrule{1-10}
 & $n = 100$ & 0.002 & 0.015 & 0.041 &  & $n = 100$ & 0.007 & 0.026 & 0.064\\
 & $n = 500$ & 0.005 & 0.018 & 0.041 &  & $n = 500$ & 0.013 & 0.056 & 0.104\\
\multirow{-3}{*}{\raggedright M4} & $n = 1000$ & 0.004 & 0.021 & 0.053 & \multirow{-3}{*}{\raggedright M9} & $n = 1000$ & 0.011 & 0.059 & 0.105\\
\cmidrule{1-10}
 & $n = 100$ & 0.009 & 0.039 & 0.091 &  & $n = 100$ & 0.009 & 0.046 & 0.085\\
 & $n = 500$ & 0.017 & 0.078 & 0.141 &  & $n = 500$ & 0.011 & 0.042 & 0.082\\
\multirow{-3}{*}{\raggedright M5} & $n = 1000$ & 0.034 & 0.104 & 0.190 & \multirow{-3}{*}{\raggedright M10} & $n = 1000$ & 0.006 & 0.034 & 0.065\\
\bottomrule
\end{tabular}
\end{table}

For the bimodal distributions (Models 6 to 10), the proportions of rejections are similar to or smaller than the significance levels in most cases. The only exception is Model 6, for which the obtained proportion of rejections at $10\%$ is larger than $0.1$ for the smallest sample size ($n = 100$). But this calibration problem is rapidly corrected increasing the sample size, and the proportions of rejections are always smaller than the signification levels for the other two sample sizes.

Comparing the results of the unimodal and bimodal distributions, the proportions of rejections are always smaller for the unimodal models than the bimodal ones. Furthermore, the only two cases where the proportions of rejections are similar to the signification levels for the largest sample size are bimodal (Models 8 and 9).

Table \ref{tab:resultbimpow} shows the power results. The new test detects satisfactorily the alternative hypothesis in the five considered scenarios. For all the models, the test obtains proportions of rejections that are larger than the significance level for medium ($n = 500$) or large ($n = 1000$) sample size, and the test power increases as $n$ grows. Comparing the results of the five models leads to some surprising conclusions. The results of the Models 13 to 15 show that the power of the new test naturally decreases as the excess mass of a secondary mode becomes smaller (Model 13 vs. Model 14) or as the two secondary modes approach each other (Models 13 vs. Model 15, see Appendix~\ref{sec:A3} for the definition of the fifteen models). Nevertheless, the test obtains smaller proportions of rejections for Model 11 than for Model 12, which, quite unexpectedly, suggests that the power of the test grows as the three modes come closer. This strange behavior may be caused by the particular distribution of Model 12, that concentrates almost all the probability between $\pi/2$ and $3\pi/2$. As previously seen with Model 5, this sectorial shape could lead to inexplicably large proportions of rejections and, therefore, be the cause of this strange phenomenon.

\begin{table}[!t]
\centering
\caption{\label{tab:resultbimpow}
		Proportions of rejections for the test based on pseudo-likelihood \eqref{eq:Dk} with significance level 1\%, 5\% and 10\% when testing bimodality. All proportions of rejections were calculated from $\mathcal{M} = 1000$ samples. $B = 500$ resamples were employed for test calibration. Each row represents a different combination of sampling distribution and sample size. All the distributions considered had $3$ modes. See Appendix~\ref{sec:A3} for a definition of the models.}
\begin{tabular}{lllll}
\toprule
Model & Size & 1\% & 5\% & 10\%\\
\midrule
 & $n = 100$ & 0.048 & 0.048 & 0.052\\

 & $n = 500$ & 0.188 & 0.189 & 0.190\\

\multirow{-3}{*}{\raggedright M11} & $n = 1000$ & 0.435 & 0.441 & 0.443\\
\cmidrule{1-5}
 & $n = 100$ & 0.037 & 0.113 & 0.183\\

 & $n = 500$ & 0.209 & 0.395 & 0.496\\

\multirow{-3}{*}{\raggedright M12} & $n = 1000$ & 0.465 & 0.643 & 0.723\\
\cmidrule{1-5}
 & $n = 100$ & 0.588 & 0.715 & 0.776\\

 & $n = 500$ & 0.999 & 0.999 & 0.999\\

\multirow{-3}{*}{\raggedright M13} & $n = 1000$ & 1.000 & 1.000 & 1.000\\
\cmidrule{1-5}
 & $n = 100$ & 0.341 & 0.530 & 0.650\\

 & $n = 500$ & 0.851 & 0.904 & 0.927\\

\multirow{-3}{*}{\raggedright M14} & $n = 1000$ & 0.959 & 0.977 & 0.983\\
\cmidrule{1-5}
 & $n = 100$ & 0.127 & 0.186 & 0.259\\

 & $n = 500$ & 0.432 & 0.584 & 0.694\\

\multirow{-3}{*}{\raggedright M15} & $n = 1000$ & 0.640 & 0.764 & 0.821\\
\bottomrule
\end{tabular}
\end{table}

\section{Real data illustration}\label{sec:data}

Once the new test calibration and power is checked and compared to the proposal by \cite{AmeijeirasAlonso2019}, the performance in practice of both tests is illustrated by analyzing the migratory bird data \citep{CabreraCruz2016} described in Section~\ref{sec:intro}. First, the post-construction observations are regarded in order to assess whether there exists more than one mode. It should be noticed that, in the simulation study performed in Section~\ref{sec:simstudy}, Model~8 was chosen to resemble the distribution of the post-construction data (see Appendix~\ref{sec:A2} for details). For the new test, the p-value is approximated with $B = 1000$ bootstrap resamples, and the estimated p-value is~$0$. For the proposal by \cite{AmeijeirasAlonso2019}, the p-value is also estimated with $B = 1000$ resamples, and the estimation is $0$ too. Therefore, both tests reject the unimodality of the data for the most common significance levels. This provides statistical evidence that migratory raptors follow at least two different strategies to avoid the two new wind farms developed in the observation area, as discussed in Section~\ref{sec:intro}.

Now, the tests are applied to the pre-construction data. In this case, the estimated p-values are $0.294$ for the new test, and $0.805$ for the test by \cite{AmeijeirasAlonso2019} (again, $B = 1000$ bootstrap resamples were used for approximating the two p-values). Consequently the two tests confirm the unimodality of the pre-construction data. Since the post-construction observations have been shown to be multimodal, this also points out that the construction of the two new wind farms caused a change in the migratory raptor routes in order to avoid them.

\section{Conclusions and discussion}\label{sec:conclusion}
This paper introduces a new multimodality test for circular data, based on likelihood ratio ideas adapted to a nonparametric context. In the exhaustive simulation study presented in Section~\ref{sec:simstudy}, the new proposal exhibits a satisfactory performance for most scenarios, showing a good calibration under the null hypothesis of unimodality (bimodality) and successfully detecting the alternative of having more than one mode (two modes). Furthermore, simulation results show that the new test is consistently more powerful than its main existing competitor, the proposal by \cite{AmeijeirasAlonso2019}, which obtains smaller rejection proportions in all the considered cases. Finally, the new test is then applied to a real dataset, illustrating the practical relevance of such procedures.

Perspectives for future research are rich and diverse. First, it could be interesting to derive the asymptotic distribution of $D_k$ defined in equation~\eqref{eq:Dk} under the null hypothesis of $\mathcal{H}_0: j \leq k$ in \eqref{eq:testmultimod}. However, the absence of a closed form expression for~$D_k$ makes it challenging to run a theoretical analysis. In addiction, as it was noted in Section~\ref{sec:newproposal}, our simulations results raise some skepticism about the utility in practice of determining the null distribution of $D_k$. Anyway, the asymptotic distribution of $D_k$ could help to understand the conservative behavior of our test under the null hypothesis.

The most promising avenue for future research consists of adapting the testing procedure to other contexts. The pseudo-likelihood approach only requires a kernel density estimator~$\hat{f}_h$ for test implementation, which has natural extensions to a large variety of spaces, such as the sphere and the hypersphere (check \citealp{Mardia2000}, Ch. 12), the torus \citep{DiMarzio2011}, or the cylinder \citep{GarciaPortugues2013a}. Then, an analogous reasoning can be followed to test multimodality for data on those spaces too.

For example, consider the case where $f$ is a density function supported on the $d$-dimensional hypersphere $\mathbb{S}^d$. Given $\bm{X}_1, \ldots, \bm{X}_n \in \mathbb{S}^d$ an i.i.d.~sample from $f$, the kernel density estimator of $f$ (with von Mises-Fisher kernel) is defined as
\begin{equation}
    \label{eq:kdesphere}
    \hat{f}_{\kappa} (\bm{x}) = a_d (\kappa) \frac{1}{n} \sum_{i = 1}^n \exp (\kappa \bm{X}_i' \bm{x}),
\end{equation}
where the quantity $\kappa \in (0, + \infty)$ controls the degree of smoothing and
\begin{equation*}
    a_d (\kappa) = \log \bigg[ \bigg(\frac{\kappa}{2} \bigg)^{1 - d/2} \Gamma \bigg( \frac{d}{2} \bigg) I_{d/2 - 1} (\kappa) \bigg].
\end{equation*}
Based on the kernel density estimator in \eqref{eq:kdesphere}, we can introduce the pseudo-likelihood function for hyperspherical data as
\begin{equation*}
	\mathcal{L}_{CV}(\kappa) = \prod_{i = 1}^{n} \hat{f}^{-i}_{\kappa} ( \bm{X}_i ), \quad h > 0;
\end{equation*}
where
\begin{equation*}
	\hat{f}^{-i}_{\kappa} (\bm{x}) = a_d (\kappa) \frac{1}{n - 1} \sum_{m = 1, m \neq i}^{n} \exp (\kappa \bm{X}_m' \bm{x}).
\end{equation*}
Then, one can address the testing problem in \eqref{eq:testmultimod} for hyperspherical data using a new version of the test statistic $D_k$ in \eqref{eq:Dk} based on this pseudo-likelihood function. As in the circular case, large values of $D_k$ reject the null hypothesis.

However, this extension is not as straightforward as it seems at first glance. For example, the new multimodality test requires that the underling density function of the data is bounded away from zero for correct calibration, something that has been already confirmed through simulations. This requirement indicates that this procedure will only exhibit a good behavior in compact manifolds (such as the sphere or the torus), since in non-compact manifolds no density can be bounded away from zero. Moreover, a key fact for the construction of the test for circular data is that the number of modes of the kernel density estimator~$\hat{f}_h$ is monotonous with $h$, a property that is not preserved in multidimensional spaces. Consequently, the division of the parameter space between the null and alternative hypotheses is not as simple as in the circular case, making the computation of~$D_k$ more challenging. The lack of monotonicity also impedes defining an analogue of the critical bandwidth in multidimensional spaces. Then, the current calibration method, which heavily relies on critical bandwidth, must be reformulated.

\begin{appendices}
\section{Proof of Proposition 1}\label{sec:A1}
	Let $\mathcal{K}_h$ the density function of a wrapped normal WN$(0, h^2)$ defined by \eqref{eq:wnorm}. It is easy to check that $\lim_{h \rightarrow + \infty } \mathcal{K}_h (x) = 1/2 \pi$ for all $x \in \mathbb R$. So:
	\begin{equation*}
		\lim_{h \rightarrow + \infty} \mathcal{L}_{CV} (h) = \lim_{h \rightarrow + \infty} \prod_{i = 1}^{n} \hat{f}^{-i}_h ( X_i ) = ( 2 \pi )^{-n}.
	\end{equation*}
	
	Split the series present in \eqref{eq:wnorm} in half, getting one series for $m \geq 0$ and other for $m < 0$. For the first series, one obtains:
	\begin{multline}
		\label{eq:ineqwn}
		\sum_{m = 0}^{+ \infty} \exp \left[ -  \frac{(x + 2 \pi m )^2}{2h^2}\right] \leq \sum_{m = 0}^{+ \infty} \exp \left( -  \frac{x^2}{2h^2} - \frac{2\pi^2 m^2}{h^2} \right) \leq \\
		\leq \exp \left( -  \frac{x^2}{2h^2}\right) \sum_{m = 0}^{+ \infty} \exp \left( - \frac{2\pi^2 m}{h^2} \right) = \exp \left( -  \frac{x^2}{2h^2} \right) \left[ 1 - \exp \left( - \frac{2\pi^2}{h^2} \right) \right]^{-1}.
	\end{multline}
	The first inequality of \eqref{eq:ineqwn} is given by $(a + b)^2 \geq a^2 + b^2$ for all $a, b \geq 0$. The last inequality is straightforward, since one is adding more terms to the series (there are more natural numbers than perfect squares).
	
	A similar reasoning shows that
	\begin{equation}
 		\label{eq:ineqwn2}
		\sum_{m = 1}^{+ \infty} \exp \left[ -  \frac{ ( x - 2 \pi m )^2}{2h^2}\right] \leq \exp \left[ -  \frac{(2 \pi - x)^2}{2h^2} \right] \left[ 1 - \exp \left( - \frac{2\pi^2}{h^2} \right) \right]^{-1}.
	\end{equation}
	With \eqref{eq:wnorm}, \eqref{eq:ineqwn} and \eqref{eq:ineqwn2}, one deduces
	\begin{equation*}
		\mathcal{K}_h ( x ) \leq \left[ 1 - \exp \left( - \frac{2\pi^2}{h^2} \right) \right]^{-1} \left\lbrace \frac{1}{\sqrt{2 \pi h^2}} \exp \left[ -  \frac{(2 \pi - x)^2}{2h^2} \right] + \frac{1}{\sqrt{2 \pi h^2}} \exp \left( -  \frac{x^2}{2h^2} \right) \right\rbrace;
	\end{equation*}
	so $\lim_{h \rightarrow 0} \mathcal{K}_h (x) = 0$ for every $x \in (0, 2 \pi)$.
	
	Assume that the sample $X_1, \ldots, X_n$ has no repeated observations, that is, $X_i \neq X_m$ for all $i \neq m$. Then,
	\begin{equation*}
		\lim_{h \rightarrow 0} \hat{f}^{-i}_h ( X_i ) = \lim_{h \rightarrow 0} \frac{1}{n - 1} \sum_{m = 1, m \neq i}^{n} \mathcal{K}_h \left( X_i - X_m \right) = 0,
	\end{equation*}
	for all $i \in \{1, \ldots, n \}$, so $\lim_{h \rightarrow 0} \mathcal{L}_{CV} (h) = 0$.
	
	Summarizing, if there are no repeated observations in $X_1, \ldots, X_n$, it has been shown that
	\begin{equation}
		\label{eq:ellCVbound}
		\lim_{h \rightarrow 0} \mathcal{L}_{CV} (h) = 0, \quad \lim_{h \rightarrow + \infty} \mathcal{L}_{CV} (h) = ( 2 \pi )^{-n}.
	\end{equation}
	$\mathcal{L}_{CV}(h)$ is a continuous function in $(0, + \infty)$. Therefore, \eqref{eq:ellCVbound} implies that $\mathcal{L}_{CV}(h)$ is bounded in $(0, + \infty)$. Since $X$ is an absolutely continuous angle, this happens with probability~1.

\section{Data and code} \label{sec:A2}
The data analyzed in this paper were collected by \cite{CabreraCruz2016a} and are available in Figshare (under license CC BY 4.0, \url{https://creativecommons.org/licenses/by/4.0/}) with the identifier: \url{https://doi.org/10.6084/m9.figshare.3123100.v2}.

The code for reproducing the simulation study can be found in \url{https://github.com/diboro/testingnmodes}. Once the manuscript will be revised, the final code will be delivered through R package NPCirc, devoted to nonparametric methods for circular data.

\section{Simulation models} \label{sec:A3}

Table~\ref{tab:models} contains the distributions of all the models that had been considered in the simulation study presented in Section \ref{sec:simstudy}. In that table, $\text{vM}(\mu, \kappa)$ denotes the von Mises distribution with mean direction $\mu$ and concentration $\kappa$ as in Section~3.5.4 of \cite{Mardia2000}, and $\mathrm{ssvM} (\mu, \kappa, \lambda)$ is the sine-skewed von Mises with skewing parameter $\lambda \in [-1, 1]$ as defined by \cite{Abe2009}.

\begin{table}[!ht]
	\centering
    \caption{\label{tab:models} The fifteen models considered in the simulation study of Section~\ref{sec:simstudy}. All models are mixtures of von~Mises except Models~4 and~5.}
	\begin{tabular}{ll}
		\toprule
		Model & Distribution \\
		\midrule
        \addlinespace[0.3em]
        \multicolumn{2}{l}{\textbf{Unimodal:}}\\
        \addlinespace[0.3em]
		\hspace{0.5em} M1 & vM$(\pi, 1)$ \\
  		\hspace{0.5em} M2 & $0.2 \cdot \text{vM}(2 \pi /3, 3) + 0.6 \cdot \text{vM}( \pi, 1.4) + 0.2 \cdot \text{vM}(4 \pi /3, 3)$ \\
		\hspace{0.5em} M3 & $0.05 \cdot \text{vM}(2 \pi /3, 7) + 0.9 \cdot \text{vM}( \pi, 1) + 0.05 \cdot \text{vM}(4 \pi /3, 7)$ \\
		\hspace{0.5em} M4 & ssvM$(\pi, 1, -0.9)$ \\
		\hspace{0.5em} M5\footnotemark[1] & $\pi \cdot \text{beta}(3,2) + \pi / 2$ \\
        \addlinespace[0.3em]
        \multicolumn{2}{l}{\textbf{Bimodal:}}\\
        \addlinespace[0.3em]
		\hspace{0.5em} M6 & $0.5 \cdot \text{vM}( \pi - 1.25, 1.5) + 0.5 \cdot \text{vM}( \pi + 1.25, 1.5)$ \\
  		\hspace{0.5em} M7 & $0.5 \cdot \text{vM}( \pi - 1, 1.5) + 0.5 \cdot \text{vM}( \pi + 1, 1.5)$ \\
		\hspace{0.5em} M8 & $0.5 \cdot \text{vM}( 1.5 , 4) + 0.5 \cdot \text{vM}( 3, 2)$ \\
		\hspace{0.5em} M9 & $0.95 \cdot \text{vM}( \pi /2, 6) + 0.05 \cdot \text{vM}( 3 \pi /2, 3)$ \\
		\hspace{0.5em} M10 & $0.9 \cdot \text{vM}( \pi /2, 6) + 0.1 \cdot \text{vM}( 3 \pi /2, 3)$ \\
        \addlinespace[0.3em]
        \multicolumn{2}{l}{\textbf{Trimodal:}}\\
        \addlinespace[0.3em]
		\hspace{0.5em} M11 & $1/3 \cdot \text{vM}(\pi-2, 7) + 1/3 \cdot \text{vM}(\pi, 7) + 1/3 \cdot \text{vM}(\pi+2, 7)$ \\
  		\hspace{0.5em} M12 & $1/3 \cdot \text{vM} (\pi-1, 7) + 1/3 \cdot \text{vM} (\pi, 7) + 1/3 \cdot \text{vM} (\pi+1, 7)$ \\
		\hspace{0.5em} M13 & $0.2 \cdot \text{vM} (\pi/2, 6) + 0.2 \cdot \text{vM} (\pi, 6) + 0.6 \cdot \text{vM} (7\pi/4, 8)$ \\
		\hspace{0.5em} M14 & $0.1 \cdot \text{vM} (\pi/2, 6) + 0.25 \cdot \text{vM} (\pi, 6) + 0.65 \cdot \text{vM} (7\pi/4, 8)$ \\
		\hspace{0.5em} M15 & $0.2 \cdot \text{vM} (\pi/2, 6) + 0.2 \cdot \text{vM} (6\pi/7, 6) + 0.6 \cdot \text{vM} (7\pi/4, 8)$ \\
		\bottomrule
	\end{tabular}
    \footnotetext[1]{Model~5 is a standard beta(3,2) distribution (as parameterized by \citealp{Johnson1995}, Ch. 25) rescaled so that its support is the interval $[\pi/2, 3\pi/2]$.}
\end{table}

\begin{figure}[!p]
    \centering
	\begin{subfigure}{0.3 \textwidth}
		\includegraphics[width = \textwidth]{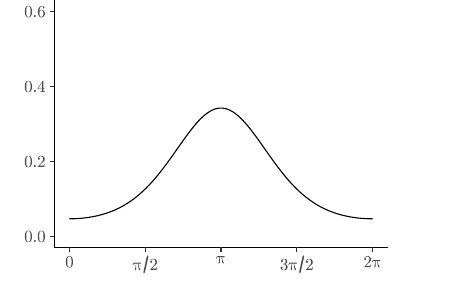}
		\caption*{Model 1.}
	\end{subfigure}
	\begin{subfigure}{0.3 \textwidth}
		\includegraphics[width = \textwidth]{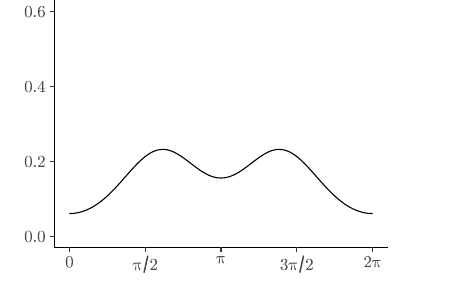}
		\caption*{Model 6.}
	\end{subfigure}
	\begin{subfigure}{0.3 \textwidth}
		\includegraphics[width = \textwidth]{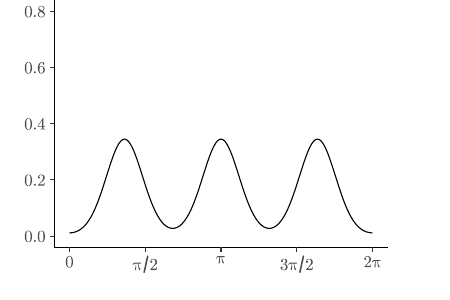}
		\caption*{Model 11.}
	\end{subfigure}
	\begin{subfigure}{0.3 \textwidth}
		\includegraphics[width = \textwidth]{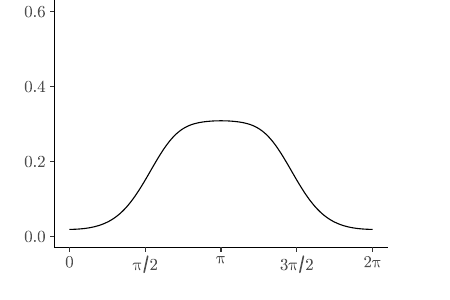}
		\caption*{Model 2.}
	\end{subfigure}
	\begin{subfigure}{0.3 \textwidth}
		\includegraphics[width = \textwidth]{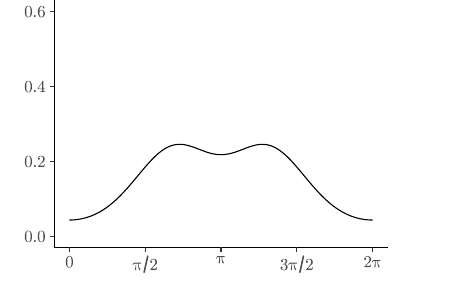}
		\caption*{Model 7.}
	\end{subfigure}
	\begin{subfigure}{0.3 \textwidth}
		\includegraphics[width = \textwidth]{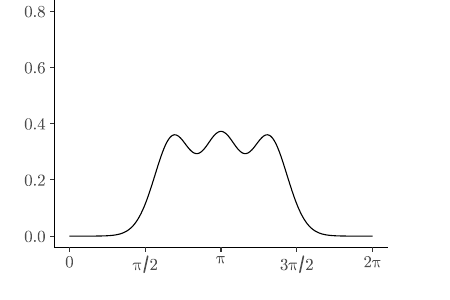}
		\caption*{Model 12.}
	\end{subfigure}
    \begin{subfigure}{0.3 \textwidth}
		\includegraphics[width = \textwidth]{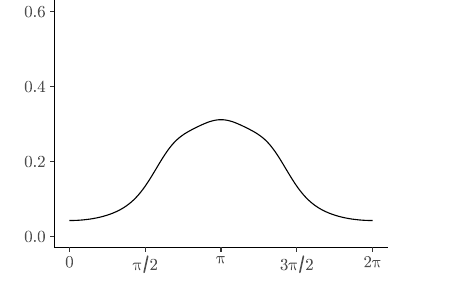}
		\caption*{Model 3.}
	\end{subfigure}
	\begin{subfigure}{0.3 \textwidth}
		\includegraphics[width = \textwidth]{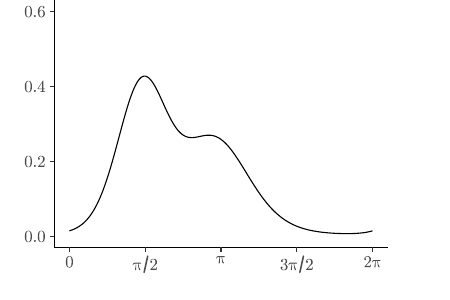}
		\caption*{Model 8.}
	\end{subfigure}
	\begin{subfigure}{0.3 \textwidth}
		\includegraphics[width = \textwidth]{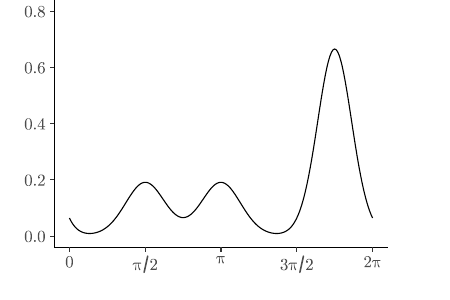}
		\caption*{Model 13.}
	\end{subfigure}
    \begin{subfigure}{0.3 \textwidth}
		\includegraphics[width = \textwidth]{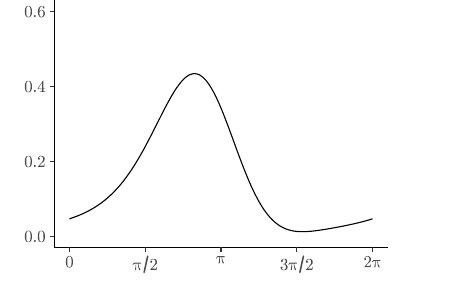}
		\caption*{Model 4.}
	\end{subfigure}
	\begin{subfigure}{0.3 \textwidth}
		\includegraphics[width = \textwidth]{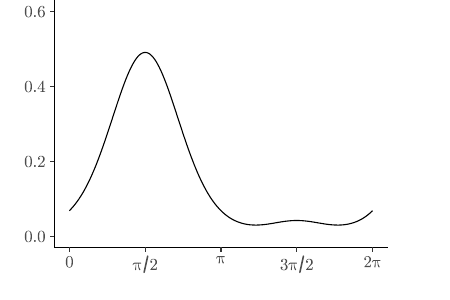}
		\caption*{Model 9.}
	\end{subfigure}
	\begin{subfigure}{0.3 \textwidth}
		\includegraphics[width = \textwidth]{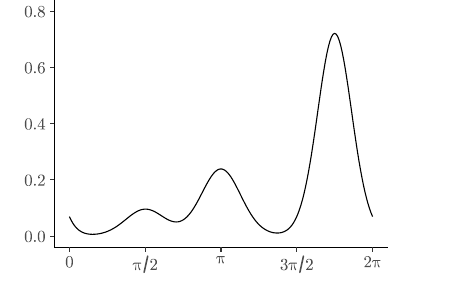}
		\caption*{Model 14.}
	\end{subfigure}
    \begin{subfigure}{0.3 \textwidth}
		\includegraphics[width = \textwidth]{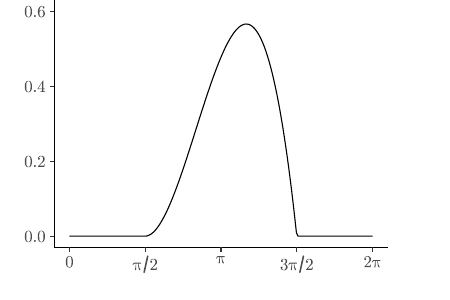}
		\caption*{Model 5.}
	\end{subfigure}
	\begin{subfigure}{0.3 \textwidth}
		\includegraphics[width = \textwidth]{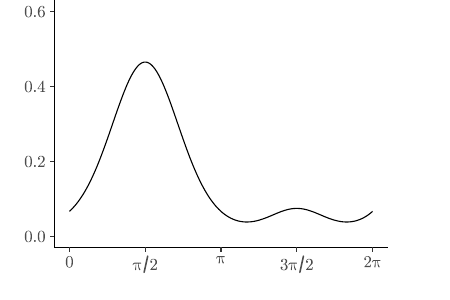}
		\caption*{Model 10.}
	\end{subfigure}
	\begin{subfigure}{0.3 \textwidth}
		\includegraphics[width = \textwidth]{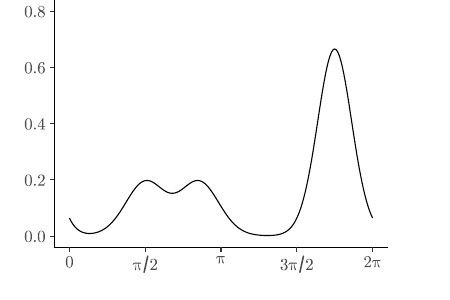}
		\caption*{Model 15.}
	\end{subfigure}
	\caption{
    \label{fig:modslin}
    Density functions of the models considered for the simulation study. Left column: unimodal models. Central column: bimodal models. Right column: trimodal models.}
\end{figure}

\begin{figure}[!p]
    \centering
	\begin{subfigure}{0.3 \textwidth}
		\includegraphics[width = \textwidth]{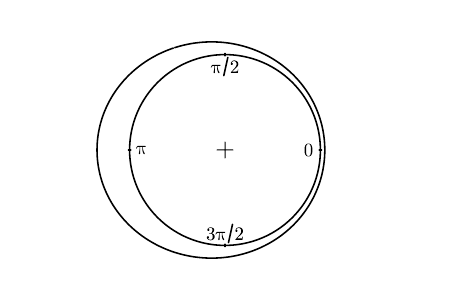}
		\caption*{Model 1.}
	\end{subfigure}
	\begin{subfigure}{0.3 \textwidth}
		\includegraphics[width = \textwidth]{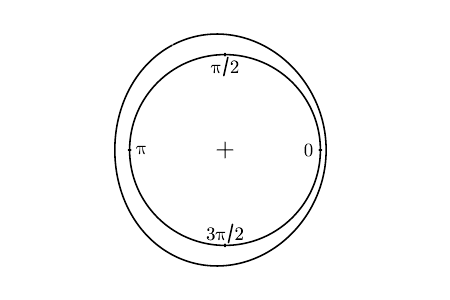}
		\caption*{Model 6.}
	\end{subfigure}
	\begin{subfigure}{0.3 \textwidth}
		\includegraphics[width = \textwidth]{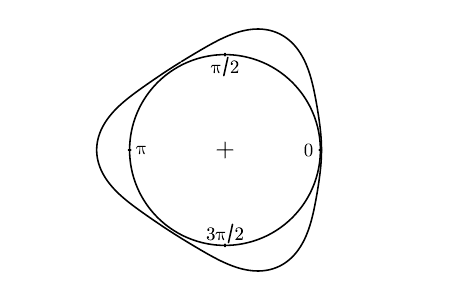}
		\caption*{Model 11.}
	\end{subfigure}
	\begin{subfigure}{0.3 \textwidth}
		\includegraphics[width = \textwidth]{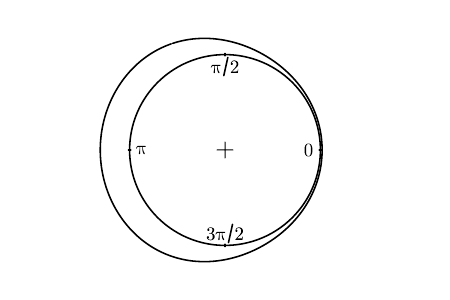}
		\caption*{Model 2.}
	\end{subfigure}
	\begin{subfigure}{0.3 \textwidth}
		\includegraphics[width = \textwidth]{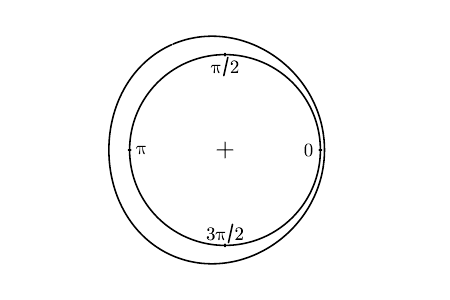}
		\caption*{Model 7.}
	\end{subfigure}
	\begin{subfigure}{0.3 \textwidth}
		\includegraphics[width = \textwidth]{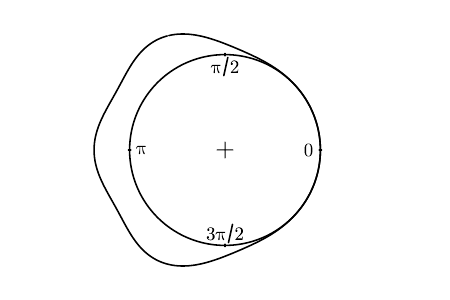}
		\caption*{Model 12.}
	\end{subfigure}
    \begin{subfigure}{0.3 \textwidth}
		\includegraphics[width = \textwidth]{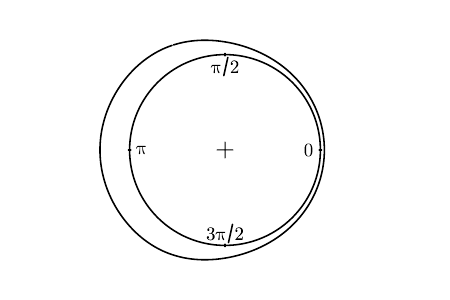}
		\caption*{Model 3.}
	\end{subfigure}
	\begin{subfigure}{0.3 \textwidth}
		\includegraphics[width = \textwidth]{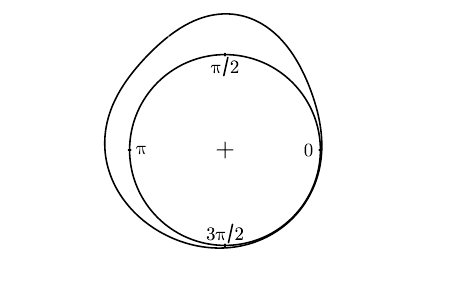}
		\caption*{Model 8.}
	\end{subfigure}
	\begin{subfigure}{0.3 \textwidth}
		\includegraphics[width = \textwidth]{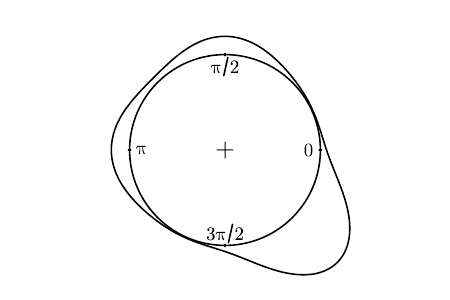}
		\caption*{Model 13.}
	\end{subfigure}
    \begin{subfigure}{0.3 \textwidth}
		\includegraphics[width = \textwidth]{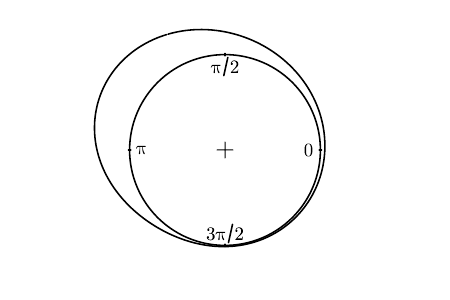}
		\caption*{Model 4.}
	\end{subfigure}
	\begin{subfigure}{0.3 \textwidth}
		\includegraphics[width = \textwidth]{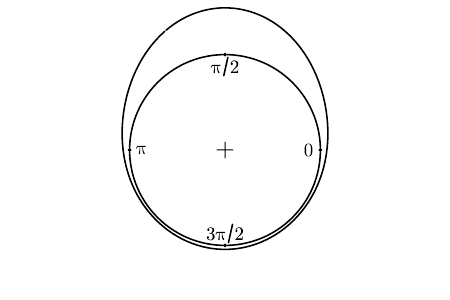}
		\caption*{Model 9.}
	\end{subfigure}
	\begin{subfigure}{0.3 \textwidth}
		\includegraphics[width = \textwidth]{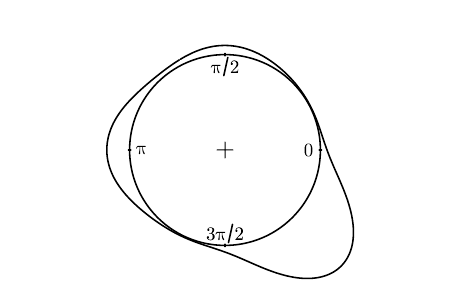}
		\caption*{Model 14.}
	\end{subfigure}
    \begin{subfigure}{0.3 \textwidth}
		\includegraphics[width = \textwidth]{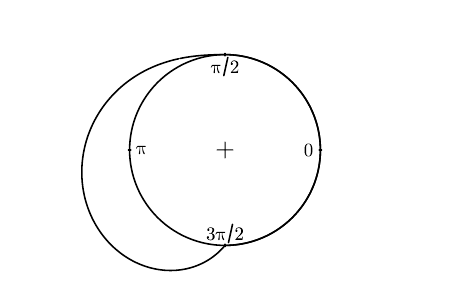}
		\caption*{Model 5.}
	\end{subfigure}
	\begin{subfigure}{0.3 \textwidth}
		\includegraphics[width = \textwidth]{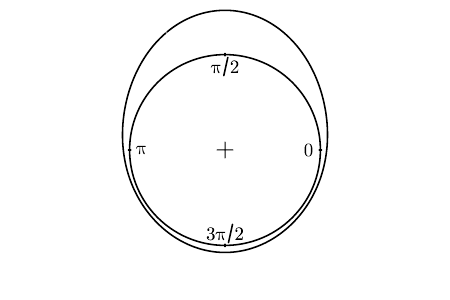}
		\caption*{Model 10.}
	\end{subfigure}
	\begin{subfigure}{0.3 \textwidth}
		\includegraphics[width = \textwidth]{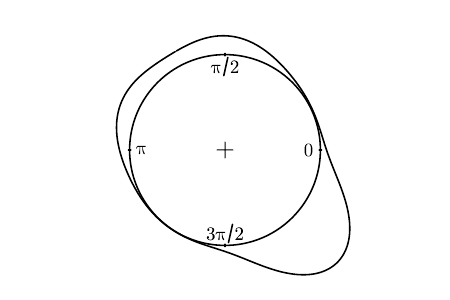}
		\caption*{Model 15.}
	\end{subfigure}
	\caption{
    \label{fig:modscirc}
    Density functions of the models considered for the simulation study. Left column: unimodal models. Central column: bimodal models. Right column: trimodal models.}
\end{figure}

Figures~\ref{fig:modslin} and~\ref{fig:modscirc} illustrate the density functions of all the fifteen models considered in the simulation study. They are represented on the real line (Figure~\ref{fig:modslin}) and on the circumference (Figure~\ref{fig:modscirc}).

\newpage

\end{appendices}

\bibliography{multimodality}

\end{document}